\documentclass[english]{iopart}
\usepackage[T1]{fontenc}
\usepackage[latin9]{inputenc}
\usepackage{geometry}
\usepackage{color}
\usepackage{array}
\usepackage{amsbsy}
\usepackage{graphicx}
\usepackage{esint}

\makeatletter

\providecommand{\tabularnewline}{\\}

\usepackage{iopams}
\usepackage{setstack}

\usepackage{tikz,pgfplots}
\usepackage{xcolor,siunitx}

\usepackage[numbers, sort&compress]{natbib}

\newcommand{\eqref}[1]{(\ref{#1})}

\makeatother

\usepackage{babel}
\begin{document}

\title[3-D modeling and simulation of 2G HTS stacks and coils]{3-D modeling and simulation of 2G HTS stacks and coils}

\author{Victor~M.~R.~Zermeño and Francesco~Grilli}

\address{{\large{Institute for Technical Physics, Karlsruhe Institute of Technology,
Germany}}}

\ead{{\large{victor.zermeno@kit.edu}}}
\begin{abstract}
Use of 2G HTS coated conductors in several power applications has
become popular in recent years. Their large current density under
high magnetic fields makes them suitable candidates for high power
capacity applications such as stacks of tapes, coils, magnets, cables
and current leads. For this reason, modeling and simulation of their
electromagnetic properties is very desirable in the design and optimization
processes. For many applications, when symmetries allow it, simple
models consisting of 1D or 2D representations are well suited for
providing a satisfying description of the problem at hand. However,
certain designs such as racetrack coils and finite-length or non-straight
stacks, do pose a 3D problem that cannot be easily reduced to a 2D
configuration. Full 3-D models have been developed, but their use
for simulating superconducting devices is a very challenging task
involving a large-scale computational problem. In this work, we present
a new method to simulate the electromagnetic transient behavior of
2G HTS stacks and coils. The method, originally used to model stacks
of straight superconducting tapes or circular coils in 2D, is now
extended to 3D. The main idea is to construct an anisotropic bulk-like
equivalent for the stack or coil, such that the geometrical layout
of the internal alternating structures of insulating, metallic, superconducting
and substrate layers is reduced while keeping the overall electromagnetic
behavior of the original device. Besides the aforementioned interest
in modeling and simulating 2G HTS coated conductors, this work gives
a further step towards efficient 3D modeling and simulation of superconducting
devices for large scale applications.
\end{abstract}


\pacs{11}

\ams{44}

\submitto{\SUST }

\maketitle

\section{Introduction}

Their high current carrying capacity at high temperatures, decreasing
price $(\$/(kA\cdot m)$) and availability in long lengths has made
of 2nd generation (2G) HTS tapes excellent candidates for several
power applications. Use of 2G HTS tapes allows compact designs for
cables such as Roebel \cite{Terzieva2010} or twisted stacked-tape
cable conductor \cite{Takayasu2012}, magnetized stacks \cite{Patel2013}
and large magnet coils \cite{Abrahamsen2010,Apte2013,Koyanagi2013}
among others. All these designs share one common structural feature:
a cross sectional plane shows a stack-like structure. For instance,
besides the case of magnetized stacks, uniformly packed coils can
be seen as a stack-like conductor making a closed loop and Roebel
cables can be modeled as a pair of stacks of tapes. 

Understanding and being able to predict the behavior of such devices
is of large importance for design and optimization purposes. Hence,
tools for modeling and simulating their electromagnetic properties
are necessary in this process. To this day, several models capable
of simulating infinitely long stacks composed of many 2G HTS tapes
are already available \cite{Rodriguez-Zermeno2011,Grilli2010,Prigozhin2011,Zermeno2013a}.
These models provide a very valuable tool for analyzing long straight
stacks and large circular coils. However, for most of the aforementioned
power applications, 3D models are needed to take into account several
factors such as the end effects in magnetized stacks or current leads,
the structure of racetrack or saddle coils or even the interaction
of long cables used in windings. Nevertheless, developments in 3D
models for these 2G HTS tapes stack-like devices is scarce and limited
to particular cases such as modeling of Roebel cables\cite{Nii2012,Zermeno2013}.
In this work, we present a homogenization tool for modeling and simulating
of these stack-like devices in 3D. 

Large circular coils and stacks made of 2G HTS tapes have already
been modeled taking into account the actual layout of the materials
in the composing tapes up to the $\rm\mu m$ scale. Examples of such
2-D simulations are found in \cite{Rodriguez-Zermeno2011} and \cite{Ainslie2011}
among others. However, computation of large scale stack-like devices
is a time demanding task. Several 2D approaches using homogenization
techniques have been proposed as a way to reduce the computing time
needed to estimate AC losses in superconducting stacks of finite height
\cite{Clem2007,Yuan2009,Prigozhin2011,Zermeno2013a}. In particular,
for large stacks modeled in 2D, use of an anisotropic homogenous-medium
approximation has provided a speedup of two orders of magnitude in
the computational time without undermining accuracy when compared
to calculations made using models describing the internal geometrical
layout \cite{Zermeno2013a}. The main goal of this work is to present
the extension of the 2D homogenization technique introduced in \cite{Zermeno2013a}
to 3D, so that more complex stack-like structures can be modeled and
simulated with ease and within a reasonable computing time.

\section{Methodology}

\begin{figure}
\begin{tabular}{>{\centering}p{1\columnwidth}}
\includegraphics[width=0.95\columnwidth]{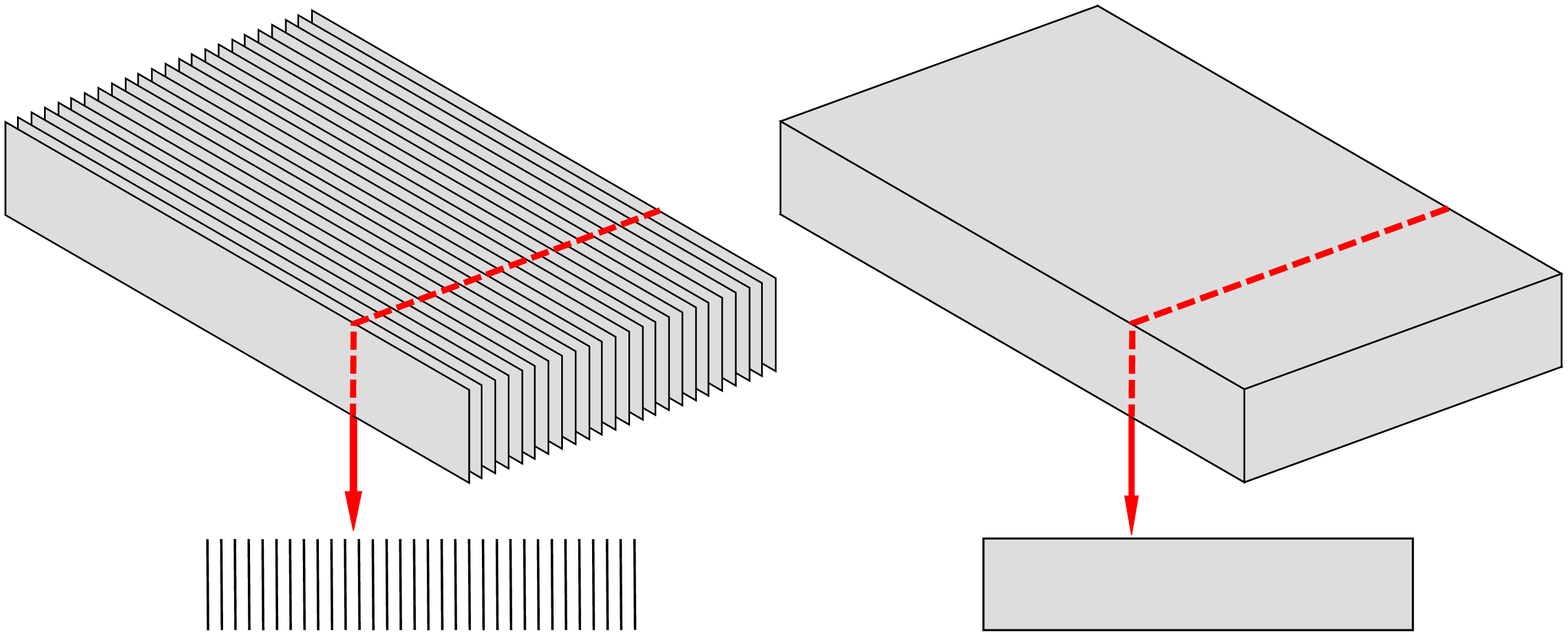}\tabularnewline
\end{tabular}

\begin{tabular}{>{\centering}p{0.5\columnwidth}||>{\centering}p{0.5\columnwidth}||>{\centering}p{0.5\columnwidth}>{\centering}p{0.47\columnwidth}}
\multicolumn{3}{>{\centering}p{0.47\columnwidth}}{$\quad\quad\quad\,\,\,\,$\includegraphics[angle=90,width=0.2625\columnwidth,height=0.065\columnwidth]{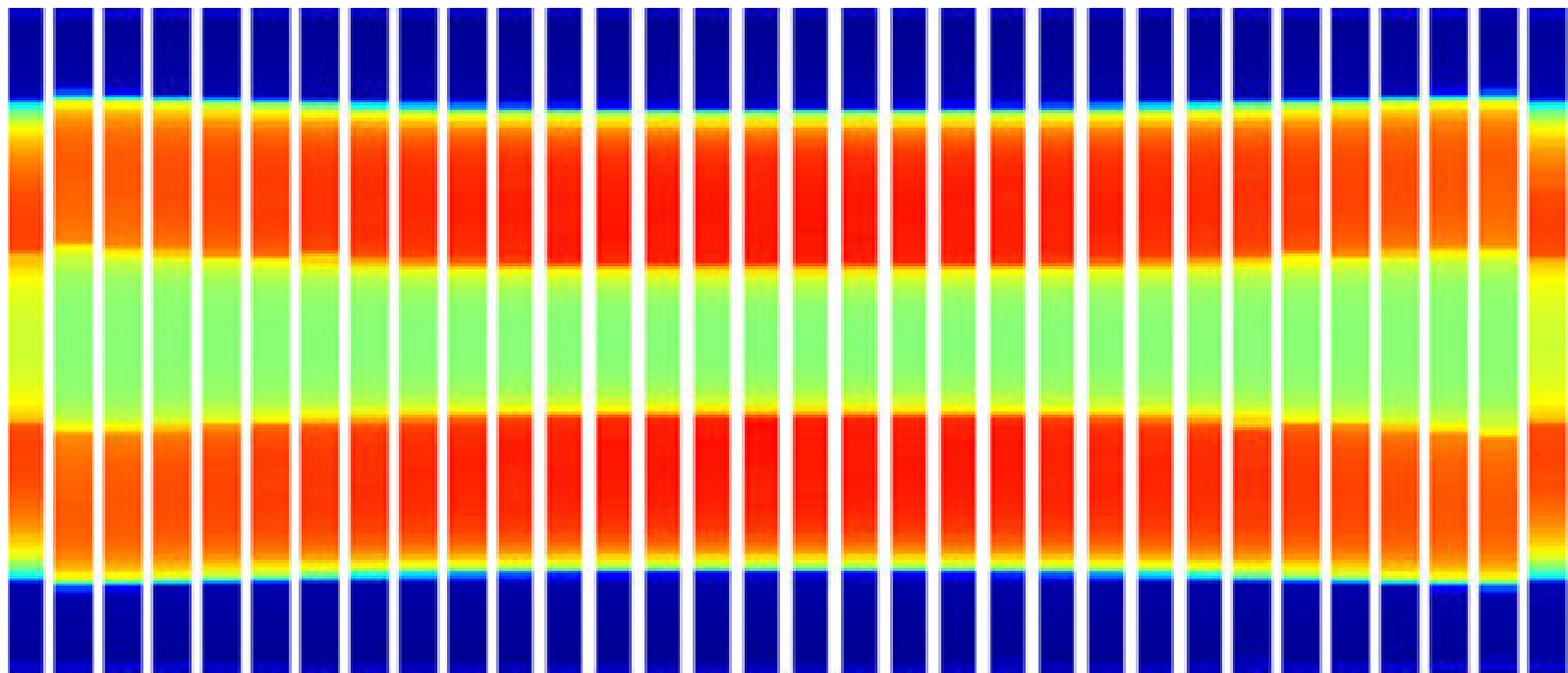}} & $\,\,\,\,\quad$\includegraphics[angle=90,width=0.2625\columnwidth,height=0.065\columnwidth]{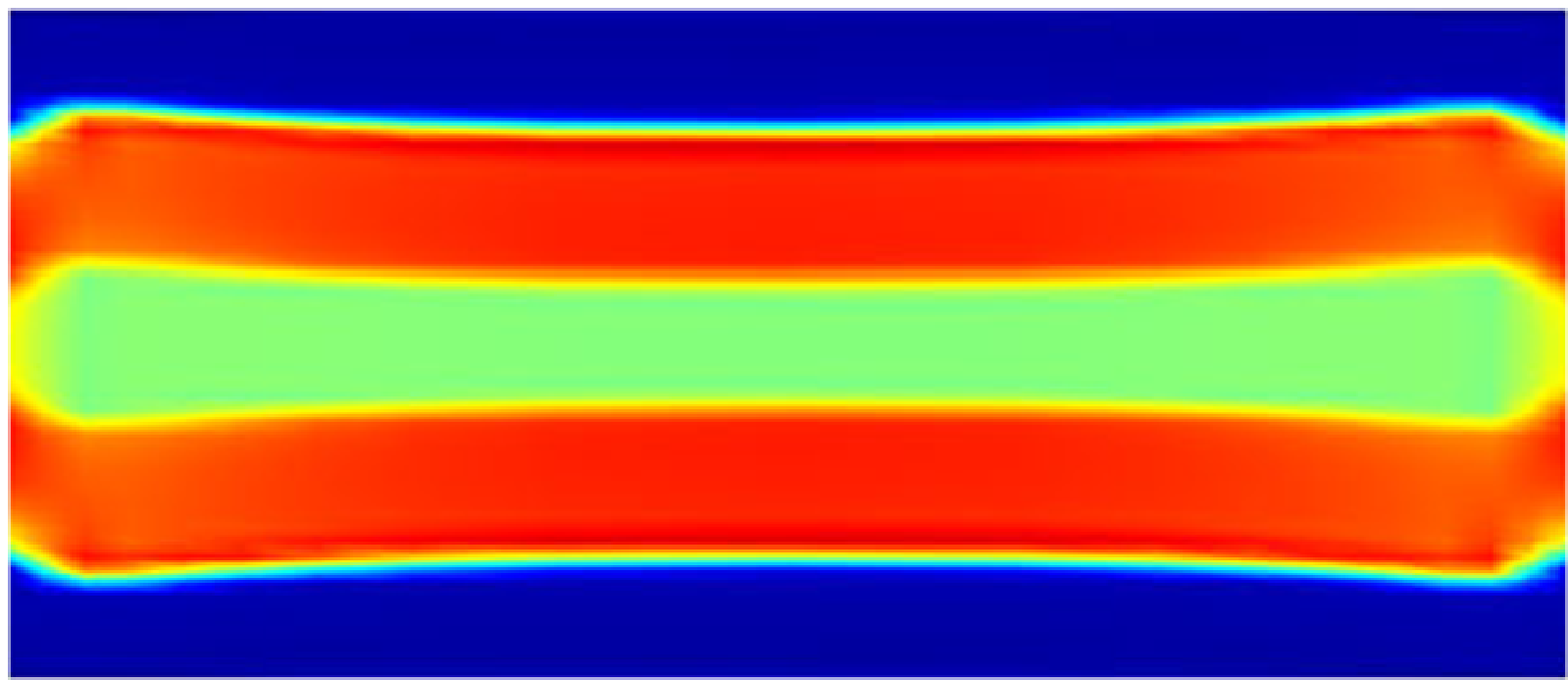}\tabularnewline
\end{tabular}

\begin{centering}
\begin{tabular}{c}
\includegraphics[scale=0.18]{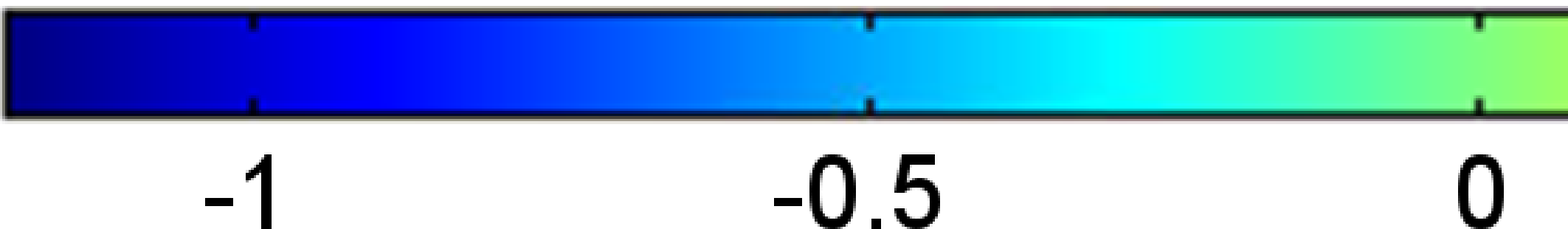}\tabularnewline
\end{tabular}
\par\end{centering}

\caption{\label{fig:concept}Comparison of a complete model considering the
actual layout of a stack of superconducting tapes (left) and its corresponding
homogenized model (right). The corresponding normalized current density
distributions are shown at the bottom. For visualization purposes,
in the model corresponding to the actual geometrical layout, the superconducting
layers\textquoteright{} actual thickness is artificially expanded
during post-processing.}
\end{figure}

Following the 2D method presented in \cite{Zermeno2013a}, in this
work we extend that approach to the 3D case. All calculations presented
here are carried out using the H-formulation of Maxwell\textquoteright{}s
equations as described in \cite{Grilli2013a} and \cite{Zermeno2013a}.
The main idea is to find a homogeneous anisotropic bulk like material
that retains the overall electromagnetic properties of a stack of
tapes carrying the same current each but with a much simpler geometrical
layout that does not include the individual tapes of the stack. The
general concept is presented in figure \ref{fig:concept} by considering
an infinite stack of tapes for the sake of simplicity. Being essentially
a 2D problem, the comparison is made using the material properties
and the method as described in \cite{Zermeno2013a}. Here, the features
on the left correspond to the actual stack where the geometrical layout
of the individual tapes is considered. The right side of the figure
presents the homogeneous bulk model. The lower part of figure \ref{fig:concept}
shows a cross-sectional view of the current distributions obtained
for both the model depicting the actual geometrical layout of the
stack and its corresponding homogeneous bulk model. As already pointed
out in \cite{Zermeno2013a}, one can note that good agreement between
both current profiles is obtained. To guarantee that the current distribution
in the homogenized stack corresponds to an equal current share in
each conductor, integral constrains are used. The following section
describes their main features and techniques regarding their implementation.

\subsection{Integral constraints}

\begin{figure}
\includegraphics[width=0.95\columnwidth]{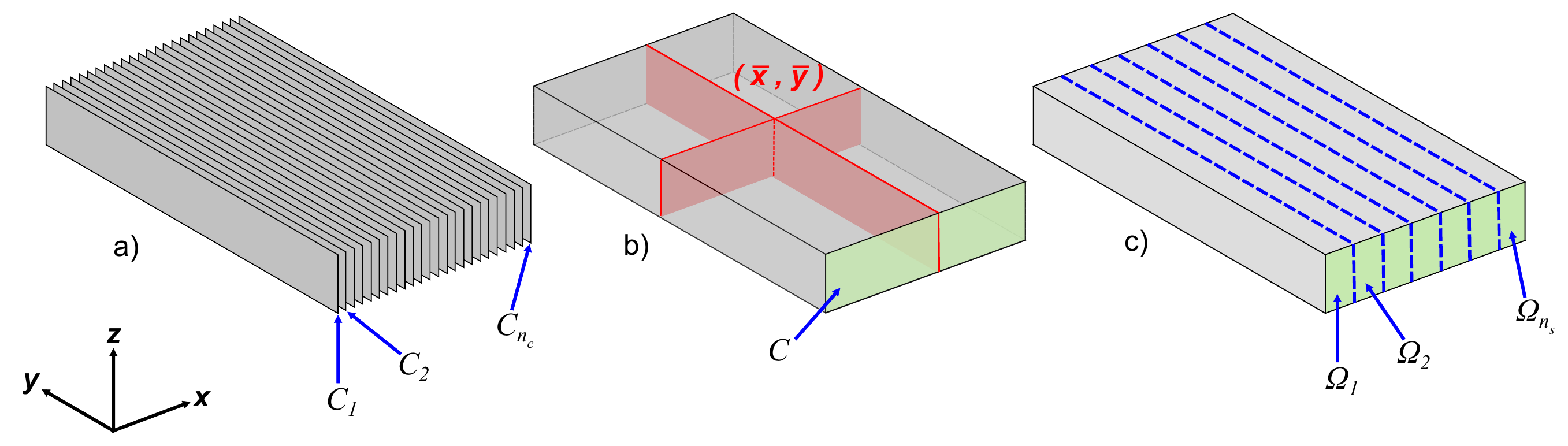}\caption{\label{fig:IntConst}Domains used for imposing the integral constraints
in a stack of tapes. Actual stack layout in which all the tapes are
considered (a), one integral constraint per conductor is needed. Homogenized
domain (b), only one integral constrain is needed. Discretized homogeneous
domain (c), one integral constraint per subdomain is needed.}
\end{figure}

When considering a stack of tapes carrying the same current, as it
is the case for the cross section of a coil, integral constraints
can be used. In that way, the net current within each conductor is
set to a given value. Considering the stack of $n_{c}$ tapes presented
in figure \ref{fig:IntConst}a, a set of integral constraints of the
form

\begin{equation}
\int_{C_{k}}\boldsymbol{J}(x,\bar{y},z,t)\cdot\hat{\boldsymbol{s_{t}}}\, dxdz=f_{k}(\bar{y},t),\,\, k\in\left\{ 1,2,\ldots,n_{c}\right\} \label{eq:IntC3D-disc}
\end{equation}
guarantees the requested current share. Here $J$ is the current density,
$f(\bar{y},t)$ is the current imposed to each tape and $\hat{\boldsymbol{s_{t}}}$
is a unitary vector defined locally as being tangential to the tapes
in the stack, pointing in the intended direction of the current flow.
One can note here that the position$\bar{y}$ where the constraint
is placed is arbitrary as the high resistivity of the air or insulation
domain will prevent the current from flowing outside the superconducting
domains making $f_{k}$ uniform. Equation (\ref{eq:IntC3D-disc})
can be rewritten as:

\begin{equation}
\int_{C_{k}}\boldsymbol{J}(x,\bar{y},z,t)\cdot\hat{\boldsymbol{s_{t}}}\, dxdz=f_{k}(t),\,\, k\in\left\{ 1,2,\ldots,n_{c}\right\} \label{eq:IntC3D-disc-1}
\end{equation}

Use of (\ref{eq:IntC3D-disc}) for the anisotropic homogenous-medium
approximation (see figure \ref{fig:IntConst}b) is not possible as
there are no separated conductors anymore that could be individually
considered. Therefore, a different approach should be used to ensure
the desired current distribution. For this purpose, one can consider
the homogenized stack as being composed of infinitely thin tapes.
Then, the following condition:

\begin{equation}
\int_{C}\boldsymbol{J}(\bar{x},\bar{y},z,t)\cdot\hat{\boldsymbol{s_{t}}}\, dz=g(\bar{x},\bar{y},t)\label{eq:IntC3D}
\end{equation}
can be used to impose a current $g(\bar{x},\bar{y},t)$ to every one
of these thin tapes. One must note again that in the case of stacks
and coils where individual tapes share the same current, $g$ becomes
uniform. Therefore, the aforementioned constraint becomes:

\begin{equation}
\int_{C}\boldsymbol{J}(\bar{x},\bar{y},z,t)\cdot\hat{\boldsymbol{s_{t}}}\, dz=g(t)\label{eq:IntC3D-1}
\end{equation}

In the following subsections three different ways to impose this condition
are presented.

\subsubsection{2D integral constraint}

\begin{figure}
\centering{}\includegraphics[width=0.5\columnwidth]{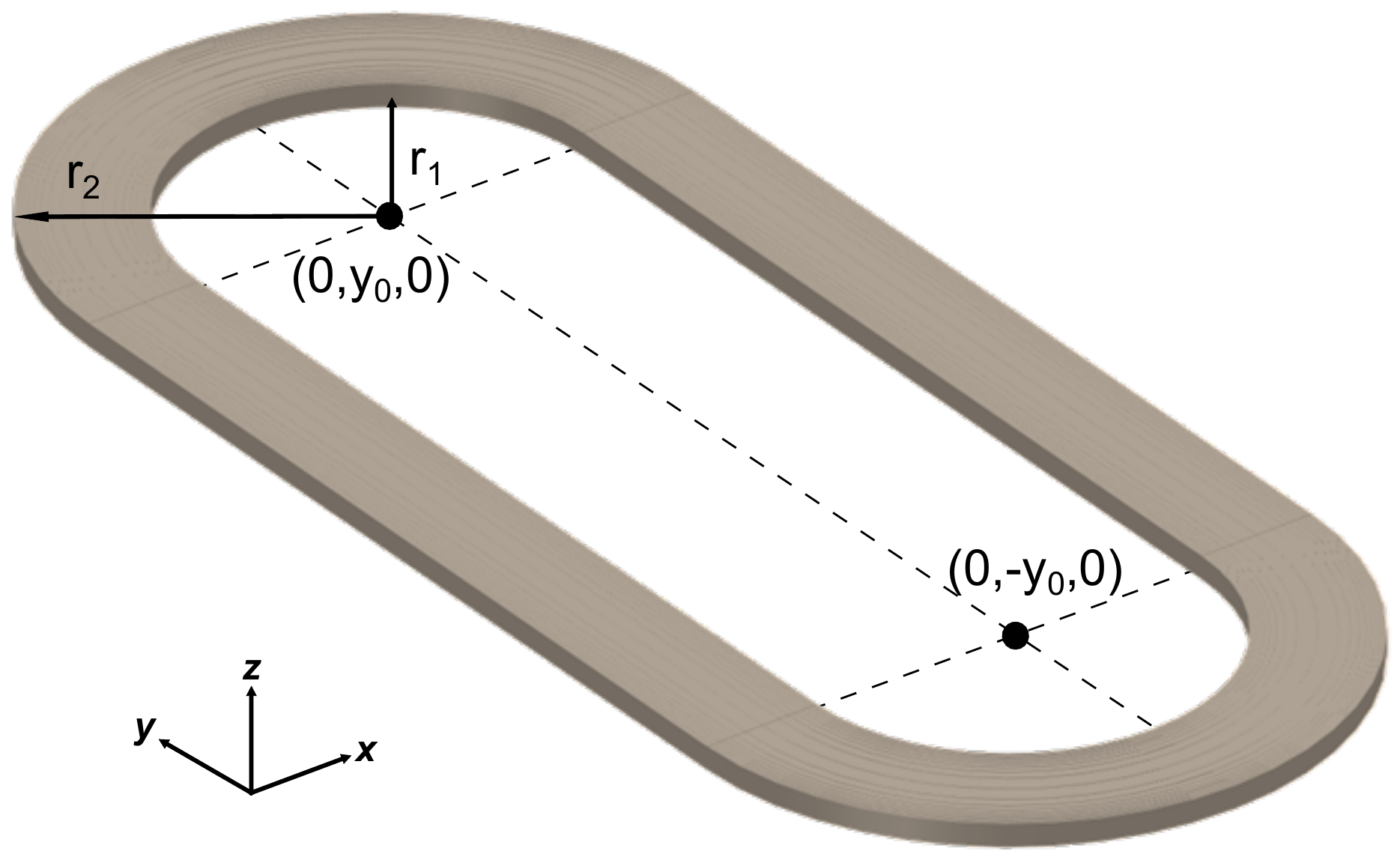}\caption{\label{fig:Racetrack}Geometrical layout of a racetrack coil.}
\end{figure}

A first approach is the direct enforcement of (\ref{eq:IntC3D-1}).
Numerically, this requires defining an integral constraint in 2D.
In this way, the constraint should specify the total current flowing
tangentially to the tapes in the stack or winding in the intended
direction of the current flow. For an infinite stack as in the one
presented in figure \ref{fig:IntConst}, this tangential direction
is parallel to the $\boldsymbol{\hat{y}}$ vector, so $\boldsymbol{\hat{s_{t}}=\boldsymbol{\hat{y}}}$.
In the case of a racetrack coil whose straight section of length $2\, y_{0}$
is aligned with the $\boldsymbol{\hat{y}}$ direction as the one in
figure \ref{fig:Racetrack}, $\boldsymbol{\hat{s_{t}}}$ is given
by the following expression:

\begin{equation}
\boldsymbol{\hat{s_{t}}}=\left\{ \begin{array}{cc}
\frac{1}{\sqrt{x^{2}+(y-y_{0})^{2}}}\left\{ -(y-y_{0}),x\right\}  & y_{0}\leq y\\
\left\{ 0,\, sign\left(x\right)\right\}  & -y_{0}\leq y<y_{0}\\
\frac{1}{\sqrt{x^{2}+(y+y_{0})^{2}}}\left\{ -(y+y_{0}),x\right\}  & y<-y_{0}
\end{array}\right.\label{eq:tang}
\end{equation}
One can note that circular planar coils can also make use of \ref{eq:tang}
by setting $y_{0}=0$. In general, expressions for other, more complicated,
windings or stack like structures can easily be found provided that
the packing of the tapes is uniform. Constraint (\ref{eq:IntC3D-1})
can then be imposed by means of Lagrange multipliers.

\subsubsection{Use of anisotropic resistivity}

One alternative procedure is to use an anisotropic resistivity tensor
that provides a very high resistivity in the direction normal to the
tapes' surface. Therefore, the $\boldsymbol{E-J}$ relationship can
be expressed as:

\begin{equation}
\boldsymbol{E_{\Vert}}=\rho_{\Vert}\boldsymbol{J}_{\Vert}\,\, and\boldsymbol{\,\, E_{\bot}}=\rho_{\bot}\boldsymbol{J}_{\bot}\label{eq:ParPerp}
\end{equation}

Here, $\rho_{\parallel}$corresponds to the homogenized resistivity
of the superconducting stack and $\rho_{\perp}$is an artificially
imposed high resistivity used to prevent current flow in the direction
normal to the tapes' surface.

\subsubsection{Manual discretization of the homogenized bulk}

Alternatively, and following a similar approach as the one presented
in \cite{Zermeno2013a}, a manual discretization approach is considered.
This implies partitioning the homogenized bulk in smaller domains
following the direction tangential to the tapes windings as shown
in figure \ref{fig:IntConst}c. The rationale behind this simplification
is that in large coils or stacks, neighboring tapes experience similar
electromagnetic conditions and hence behave in an comparable manner.
To prevent current sharing among the subdomains, highly resistive
layers are placed between them. Then the subdomains $\Omega_{1},\Omega_{2},\ldots\Omega_{n_{s}},$
of the homogeneous bulk correspond to a bundle of neighboring tapes.
Hence, by using only one element to discretize the thickness of the
subdomain, a similar behavior for each tape in the bundle is ensured.
Computationally this has a similar effect to the strategies discussed
in the previous sections, but allows for a finer control upon the
mesh density, the number of degrees of freedom and ultimately the
computational time required for simulations.

\subsection{$\boldsymbol{J_c(B)}$ dependence and $\boldsymbol{E-J}$ relationship}

As mentioned before, in this work, all calculations are made using
the H-formulation of Maxwell\textquoteright{}s equations \cite{Grilli2013a}.
To account for the magnetic flux density dependence of the critical
current density $J_{c}$, the following expression is used:

\begin{equation}
J_{c}\left(B_{\parallel},B_{\perp}\right)=\frac{J_{c_{0}}\, f_{HTS}}{\left[1+\sqrt{\left(B_{\parallel}k\right)^{2}+B_{\perp}^{2}}\Big/B_{c}\right]^{b}}\label{eq:JcB}
\end{equation}
here $B_{\parallel}$and $B_{\perp}$are respectively, the parallel
and perpendicular components of the magnetic flux density. The parameters
$J_{c_{0}},k,B_{c}$ and $b$ have the respective values of $49\,{\rm GA/m^{2}},0.275,32.5\,{\rm mT}$
and $0.6$. These parameters correspond to a characterization (not
reported here) for the $J_{c}\left(B_{\parallel},B_{\perp}\right)$
for the HTS tape used in \cite{Kario2013} using an elliptic fit.
This corresponds to a critical current $I_{c}$ of 160 A . The remaining
parameter, $f_{HTS}$, is the volume fraction of the superconducting
material in the homogenized bulk\cite{Zermeno2013a}. In what follows,
we will refer to $J_{c}\left(B_{\parallel},B_{\perp}\right)$ simply
as $J_{c}\left(\boldsymbol{B}\right)$. To describe the $\boldsymbol{E-J}$
relationship, a power law is used:

\begin{equation}
\boldsymbol{E}=E_{0}\left|\frac{J}{J_{c}\left(B_{\parallel},B_{\perp}\right)}\right|^{n}\frac{\boldsymbol{J}}{\left|J\right|}\label{eq:PL}
\end{equation}
here, the critical electrical field $E_{0}$ at which $J=J_{c}\left(B_{\parallel},B_{\perp}\right)$
is set equal to$1{\rm \,\mu V/cm}$. The exponent $n=21$ in the power
law is used to describe how abrupt is the transition from the superconducting
to normal state.

\section{Results}

In this section first we present a validation test case for the proposed
homogenization method, then we apply it to simulate a typical 3D case:
racetrack coils.

\subsection{Test case for validation}

\begin{figure}
\begin{centering}
\begin{tikzpicture}
	\pgfplotsset{every axis legend/.append style={at={(0.3,0.75)},anchor=south}}
	\begin{loglogaxis}[clip marker paths=true,
		xlabel=Current  (\si{\ampere}),
		ylabel=AC loss (\si{\joule/\meter/cycle}),
		xmin=15,xmax=120,
		xtick={20,40,60,80,100},
		xticklabels={20,40,60,80,100},
		ymin=0.001,ymax=1,
		grid=both
	] 		
		\addplot [sharp plot,mark=x,black,dashed] table[x expr=\thisrowno{0},y expr=(\thisrowno{1})]	
			{verify.txt}; 		
		\addplot [sharp plot,mark=o,black] table[x expr=\thisrowno{0},y expr=(\thisrowno{2})]
			{verify.txt}; 		
		\legend{2D Original, 3D Homogenized}
	\end{loglogaxis}
\end{tikzpicture}
\par\end{centering}

\centering{}\caption{\label{fig:Verification}Comparison of the AC loss in a stack by modeling
the original stack in 2D and as a 3D anisotropic homogenous-medium
approximation.}
\end{figure}
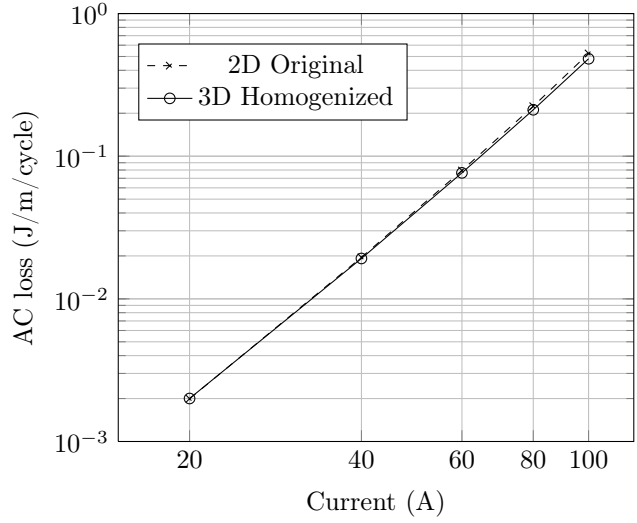

To validate the proposed strategy, the case of a stack consisting
of 50 4mm-wide tapes having a total height of 2 cm was considered.
Considering a thickness for the superconducting layer of each tape
of $1\,{\rm \mu m}$, this corresponds to a volume fraction, $f_{HTS}$
of $2.5\times10^{-3}$. AC currents at 50 Hz were imposed in each
of the tapes in the stack. The straight stack was modeled following
two different approaches: a 2D model for large stacks as described
in \cite{Rodriguez-Zermeno2011} which takes into account the 50 different
conductors in the original layout of the stack, and the 3D homogenized
model as described in the previous section. Since the conductivity
of the superconducting layers is much higher than that of the other
materials involved, and no magnetic substrate is used, only the superconducting
layers were considered. As a mean to compare both models, AC losses
were computed for different current amplitudes. The results are shown
in figure \ref{fig:Verification}. It is important to note the good
agreement over the large range of applied currents and over more than
two orders of magnitude for the calculated AC losses.

\subsection{Modeling and simulation of racetrack coils}

\begin{figure}
\centering{}\includegraphics[width=0.5\columnwidth]{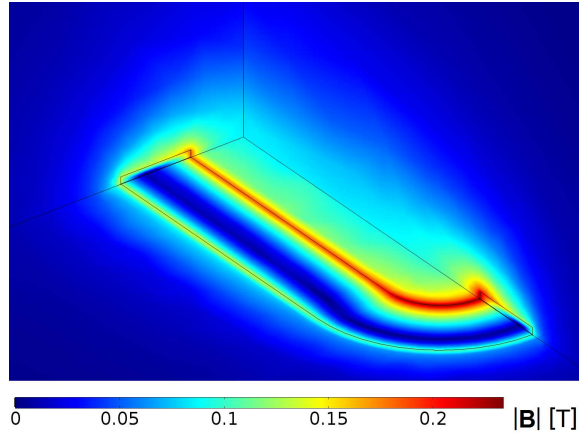}\caption{\label{fig:cross-sections}Magnitude of the magnetic flux density
in the three symmetry planes of the racetrack coil at a peak current
value of 100 A. }
\end{figure}

Having successfully tested the proposed 3D homogenization with a 2D
stack, the aim is set in addressing more complicated problems. Just
like infinitely long straight stacks can me modeled in 2D, circular
coils can be modeled using an asymmetrical approach, yielding again
a 2D model. Therefore, we have chosen to use the proposed homogenization
technique for a design that can not be modeled with 2D tools: racetrack
coils. Racetrack coils, as the one shown in figure \ref{fig:Racetrack}
have two straight and 2 round sections.Taking advantage of the symmetry
planes, only one eighth of the coil was considered for modeling and
simulation. Material properties used for simulating the racetrack
coil were the same as the ones described for the stack in the previous
section. The model considered a 50 turn single racetrack coil wound
using $4\,{\rm mm}$ wide tape with geometrical parameters $y_{0}=7.5\,{\rm cm},\, r_{1}=1.5\,{\rm cm}$
and $r_{2}=3.5\,{\rm cm}$ as shown in figure\ref{fig:Racetrack}.
Just like before, a volume fraction, $f_{HTS}$ of $2.5\times10^{-3}$was
considered. AC currents at 50 Hz were imposed to the homogenized coil. 

figure \ref{fig:cross-sections} shows the magnitude of the magnetic
flux density at the 3 symmetry planes of the racetrack coil at a peak
current value $I$ of $100\,{\rm A}$. The highest magnetic flux density
is localized in the inner part of the round section, with a value
of $234\,{\rm mT}$. In the internal region of the coil, close to
the coil's plane, the magnitude of the field is almost uniform with
a value of about $10\,{\rm mT}$. However, clear differences are observed
in both the straight and the round sections of the coil, showing the
true three dimensional structure of the model.

\begin{figure}
\includegraphics[width=0.95\columnwidth]{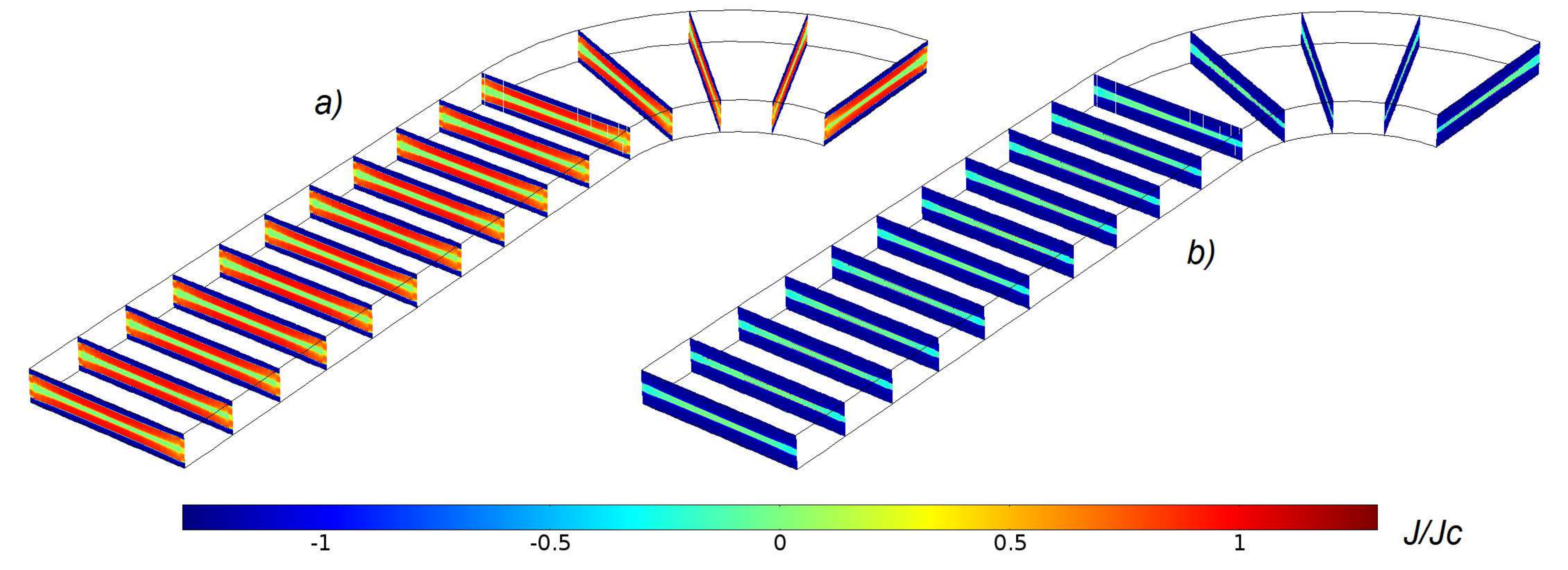}\caption{\label{fig:JoJc}Normalized current density $J/J_{c}\left(\boldsymbol{B}\right)$
at zero crossing (a) and at peak current (b).}
\end{figure}

\begin{figure}
\centering{}\includegraphics[bb=0bp 0bp 1217bp 479bp,width=1\paperwidth]{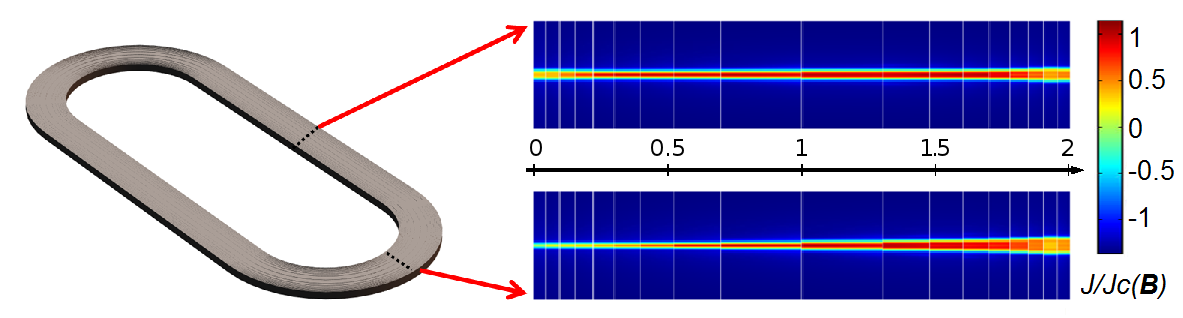}\caption{\label{fig:old_limiting-turn}Cross section plot of the normalized
current density $J/J_{c}\left(\boldsymbol{B}\right)$ at peak current
for the middle of the straight section (top) and the middle of the
round section (bottom). The rule in the center of the figure shows
the distance (in cm) from the innermost turn of the coil.}
\end{figure}

Normalized current density distributions $J/J_{c}\left(\boldsymbol{B}\right)$
for both zero crossing (a) and peak current (b) values are shown in
figure \ref{fig:JoJc} for the case of a transport current of $100\,{\rm A}$.
One can note how homogenization method allows for calculating the
current distribution in any place within the stacked tapes. At a first
glance, it might seem that the current distributions for each particular
case (a) and (b) are rather uniform in the direction of the coil's
winding. However as it will be shown that is not necessarily the case.
figure \ref{fig:old_limiting-turn} shows the current distributions
close to peak value for a transport current of $120\,{\rm A}$ for
two particular locations: the middle of the straight section (top
right) and the middle of the round section (bottom right). First,
one can note that the current distributions are not symmetric, they
have larger critical regions close to the coil's inner side. Secondly,
this effect is considerable bigger in the cross section corresponding
to the round part. This means that the current is limited by the innermost
turn of the coil in the round section. Looking at figure \ref{fig:cross-sections}
it is easy to see how this relates to the high magnetic flux density
in this region of the coil and the fact that an elliptic $J_{c}(\boldsymbol{B})$
relationship was used.

The model presented here has proved useful to calculate magnetic field,
current distribution and critical current for a coil under current
transport conditions. All of this within a 3D framework. 

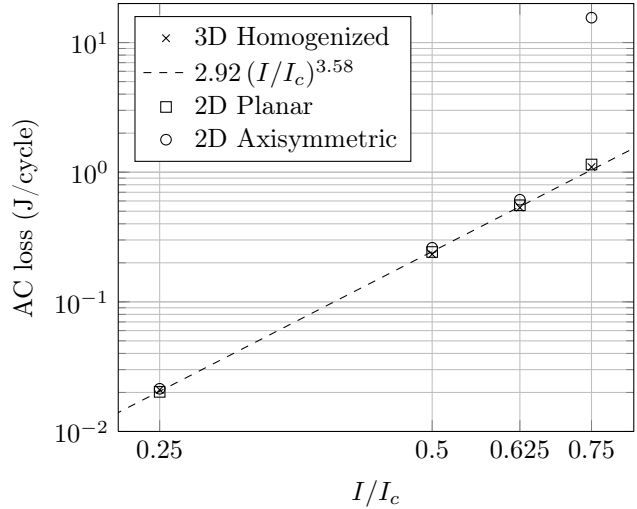
\begin{figure}
\begin{centering}
\begin{tikzpicture}
	\pgfplotsset{every axis legend/.append style={at={(0.3,0.63)},anchor=south}}
	\begin{loglogaxis}[clip marker paths=true,
		xlabel=$I/I_c$,
		ylabel=AC loss (\si{\joule/cycle}),
		xmin=0.225,xmax=0.835,
		ymin=1e-2,ymax=20,
		xtick={0.25,0.5,0.625,0.75},
		xticklabels={0.25,0.5,0.625,0.75},
		grid=both,
		legend cell align=left
	] 		
		\addplot [only marks,mark=x,black] table[skip first n=2,x expr=\thisrowno{0}/160,y expr=(\thisrowno{1})]	
			{Racetrack.txt};
		\addplot[sharp plot,mark=none,black,dashed] plot coordinates {(0.15,2.92*0.15^3.58)(0.25,2.92*0.25^3.58)(0.5,2.92*0.5^3.58)(0.625,2.92*0.625^3.58)(0.75,2.92*0.75^3.58)(0.85,2.92*0.85^3.58)};
		\addplot [only marks,mark=square,black] table[skip first n=2,x expr=\thisrowno{0}/160,y expr=(\thisrowno{2})]	 			{Racetrack.txt};
		\addplot [only marks,mark=o,black] table[skip first n=2,x expr=\thisrowno{0}/160,y expr=(\thisrowno{3})]	 			{Racetrack.txt}; 
		\legend{3D Homogenized, $2.92\,(I/I_c)^{3.58}$	,2D Planar,	2D Axisymmetric}
	\end{loglogaxis}
\end{tikzpicture}

\par\end{centering}

\centering{}\caption{\label{fig:ACloss racetrack}Computed AC losses with three different
models: 3D homogenized, 2D planar and 2D axisymmetric. A power law
fit of the form $Q=2.92\,(I/Ic)^{3.58}$ for the 3D homogenized model
is also shown (dashed line)}
\end{figure}

Finally, AC losses were computed for different current amplitudes
at 50 Hz for a 50 turn coil made with geometrical parameters $y_{0}=7.5\,{\rm cm}$,$\, r_{1}=3.5\,{\rm cm}$
and $r_{2}=5.5\,{\rm cm}.$ These correspond to the coil frame designed
as part of the Superwind project at DTU \cite{Abrahamsen20111464}
. These estimates are presented in figure \ref{fig:ACloss racetrack}.
The dashed line shows a power law fit of the form $Q=2.92\,(I/I_{c})^{3.58}$.
Here, $Q$ is the AC loss in \si{\joule/cycle}, $I$ is the amplitude
of the current and $I_{c}$ the self field critical current of the
tape (160 \si{\ampere}). A coil of similar size was studied in \cite{Rodriguez-Zermeno2011}
by means of a 2D planar model. Although, the tape characteristics
in \cite{Rodriguez-Zermeno2011} were different, in said work a similar
exponent (3.6351) in the power law fit is reported. For the purpose
of comparison, losses were also calculated using both 2D planar and
2D axisymmetric models. The 2D planar model assumed two opposing stacks
of 50 tapes each separated by $7\,{\rm cm}$. Losses per unit volume
were computed and scaled to the volume of the racetrack coil. The
2D axisymmetric model assumed a circular coil of 50 tapes with inner
radius of $3.5\,{\rm cm}$ and outer radius of $5.5\,{\rm cm}$. Again,
losses per unit volume were computed and scaled to the volume of the
racetrack coil

As shown in figure \ref{fig:ACloss racetrack}, losses computed using
a 2D axisymmetric model provided good agreement only for a limited
range of currents but diverged for large $I/I_{c}$ values. This behavior
can be easily understood since for a given current amplitude, the
field in the innermost turn of a circular coil is larger than that
of its corresponding racetrack coil. Therefore, as the current increases,
the innermost turn of the circular coil will saturate faster than
that of the racetrack coil, hence yielding much larger losses. On
the other hand, the 2D planar model showed good agreement with the
3D homogenization method. For the reasons presented above, this agreement
is expected to decrease for larger currents when the effect of the
round section of the racetrack coil becomes more important or for
coils with smaller straight sections.

\section{Conclusions}

In this work, a 3D homogenization technique used to model stacks and
coils of 2G HTS coated conductors was presented. The method complements
previous work of ours where the 2D case is addressed. Three different
ways to impose the integral constraint condition can be chosen and
for ease in implementation and computation, a strategy based upon
manual discretization of the homogenized bulk was adopted here. Although
no zero conductivity perpendicular to the tapes\textquoteright{} surface
was implemented, to prevent current sharing among the subdomains,
highly resistive layers were placed between them, proving to be a
good solution. 

For validation purposes, the 3D homogenization method was tested against
a state of the art 2D model considering all the individual conductors
and enforcing all the individual currents. Both methods provided a
remarkable good agreement over a large range of applied currents and
over more than two orders of magnitude for the calculated AC losses.

Finally, racetrack coils which provide a more complicated layout that
can not be accurately modeled using 2D methods were considered. The
3D homogenization model allowed to investigate the magnetic field
and the current distributions within the coils for the case of transport
current. The method was latter used to estimate AC losses in a racetrack
coil under transport current at several amplitudes. For comparison
AC losses were also computed using a 2D axisymmetric model, and a
2D planar model. As expected, all models converged in the low current
amplitude regime, but the 2D axisymmetric model diverged for large
$I/I_{c}$ values. Remarkably, the 2D planar model showed good agreement
with the 3D homogenized model for all the currents studied. However,
as explained in the previous section, this agreement is expected to
decrease for larger current amplitudes.

Furthermore, the 3D homogenization technique presented here is useful
for determining not just the value of the critical current but also
its distribution within the coil. This makes this homogenized model
a valuable tool for further coil optimization and for considering
interactions with other coils or materials and even as part of a larger
rotating machinery, where more complicated effects that can not be
modeled in 2D take place close to the machine's ends. \textcolor{black}{Other
areas of application for this work include transport current studies
of non planar coils and other stack-like structures such as cables.
}Future work will involve testing this model experimentally.

\ack{}{}

This work was funded partly by the Helmholtz-University Young Investigator
Group Grant VH-NG-617. The authors would like to acknowledge Dr. Nenad
Mijatovic (Department of Electrical Engineering at the Technical University
of Denmark) for his technical advice regarding the use of an anisotropic
resistivity to enforce the current flow in the direction tangential
to the tapes' winding. V. Zermeno would like to thank Dr Mads P. Soerensen
(Department of Applied Mathematics at the Technical University of
Denmark) for access to his computational resources.

\section*{-----------------}

\bibliographystyle{unsrt}
\bibliography{harvard,My_Collection_07-10-2013}

\end{document}